\documentclass[twocolumn,aip,prd,10pt,superscriptaddress,longbibliography,floatfix,citeautoscript]{revtex4-2}

\usepackage{booktabs}
\usepackage{amssymb}
\usepackage{amsmath}
\usepackage{multirow}
\usepackage{diagbox}
\usepackage{glossaries}
\usepackage{graphicx}% Include figure files
\usepackage{dcolumn}% Align table columns on decimal point
\usepackage{bm}% bold math
\usepackage{subfigure}
\usepackage[usenames,dvipsnames,svgnames]{xcolor}
\usepackage{hyperref}
\hypersetup{
pdfnewwindow=true,      % links in new window
colorlinks=true,        % false: boxed links; true: colored links
linkcolor=Blue,         % color of internal links
citecolor=Blue,         % color of links to bibliography
filecolor=Blue,         % color of file links
urlcolor=Blue           % color of external links
}
\usepackage{listings} % command insert
\usepackage{siunitx}

\newacronym{md}{MD}{molecular dynamics}
\newacronym{emd}{EMD}{equilibrium molecular dynamics}
\newacronym{nemd}{NEMD}{non-equilibrium molecular dynamics}
\newacronym{hnemd}{HNEMD}{homogeneous non-equilibrium molecular dynamics}
\newacronym{mlp}{MLP}{machine learned potential}
\newacronym{dft}{DFT}{density functional theory}
\newacronym{aimd}{AIMD}{\emph{ab initio} molecular dynamics}
\newacronym{nep}{NEP}{neuroevolution potential}
\newacronym{asi}{a-Si}{amorphous silicon} 
\newacronym{csi}{c-Si}{cubic silicon}
\newacronym{cge}{c-Ge}{cubic germanium}
\newacronym{cnt}{CNT}{carbon nanotube}
\newacronym{nhc}{NHC}{Nos{\'e}-Hoover chain}
\newacronym{rmse}{RMSE}{root-mean-square error}
\newacronym{mlmd}{MLMD}{MD simulation driven by MLP}
\newacronym{ltc}{LTC}{lattice thermal conductivity}
\newacronym{dp}{DP}{deep potential}

\begin{document}
\title{Correcting force error-induced underestimation of lattice thermal conductivity in machine learning molecular dynamics}
\author{Xiguang Wu}
\thanks{These authors contributed equally to this work.}
\affiliation{Guangzhou Key Laboratory of Low-Dimensional Materials and Energy Storage Devices, School of Materials and Energy, Guangdong University of Technology, Guangzhou 510006, China}
%\email{xgwu329@163.com}

\author{Wenjiang Zhou}
\thanks{These authors contributed equally to this work.}
\affiliation{Department of Energy and Resources Engineering, Peking University, Beijing 100871, China}
\affiliation{School of Advanced Engineering, Great Bay University, Dongguan 523000, China}
%\email{wjzhou@stu.pku.edu.cn}

\author{Haikuan Dong}
%\email{donghaikuan@163.com}
\affiliation{College of Physical Science and Technology, Bohai University, Jinzhou 121013, China}

\author{Penghua Ying}
%\email{hityingph@163.com}
\affiliation{Department of Physical Chemistry, School of Chemistry, Tel Aviv University, Tel Aviv, 6997801, Israel}

\author{Yanzhou Wang}
\affiliation{MSP group, QTF Centre of Excellence, Department of Applied Physics, Aalto University, FI-00076 Aalto, Espoo, Finland}
%\email{yanzhowang@gmail.com}

\author{Bai Song}
\email{songbai@pku.edu.cn}
\affiliation{Department of Energy and Resources Engineering, Peking University, Beijing 100871, China}
\affiliation{Department of Advanced Manufacturing and Robotics, Peking University, Beijing 100871, China}
\affiliation{National Key Laboratory of Advanced MicroNanoManufacture Technology, Beijing, 100871, China}

\author{Zheyong Fan}
\email{brucenju@gmail.com}
\affiliation{College of Physical Science and Technology, Bohai University, Jinzhou 121013, China}

\author{Shiyun Xiong}
\email{syxiong@gdut.edu.cn}
\affiliation{Guangzhou Key Laboratory of Low-Dimensional Materials and Energy Storage Devices, School of Materials and Energy, Guangdong University of Technology, Guangzhou 510006, China}

\begin{abstract}
Machine learned potentials (MLPs) have been widely employed in molecular dynamics (MD) simulations to study thermal transport. However, literature results indicate that MLPs generally underestimate the lattice thermal conductivity (LTC) of typical solids. Here, we quantitatively analyze this underestimation in the context of the neuroevolution potential (NEP), which is a representative MLP that balances efficiency and accuracy. Taking crystalline silicon, GaAs, graphene, and PbTe as examples, we reveal that the fitting errors in the machine-learned forces against the reference ones are responsible for the underestimated LTC as they constitute external perturbations to the interatomic forces. Since the force errors of a NEP model and the random forces in the Langevin thermostat both follow a Gaussian distribution, we propose an approach to correcting the LTC by intentionally introducing different levels of force noises via the Langevin thermostat and then extrapolating to the limit of zero force error. Excellent agreement with experiments is obtained by using this correction for all the prototypical materials over a wide range of temperatures. Based on spectral analyses, we find that the LTC underestimation mainly arises from increased phonon scatterings in the low-frequency region caused by the random force errors. 
\end{abstract}
\maketitle

\section{Introduction}
\Gls{ltc} of solids is a crucial physical property in many applications including thermal management of electronics \cite{MOORE2014merging, kang2021integration}, thermoelectric energy conversion \cite{snyder2008complex,zhao2014ultralow,Zhou2022effect}, and thermal barrier coatings \cite{Perepezko2009the,Robert2010overview}. Predicting and engineering \gls{ltc} \cite{qian2021Phonon-engineered} is therefore of broad interest. Nevertheless, challenges abound owing to the presence of complex structures \cite{Tadano2018quartic}, defects \cite{Hanus2021thermal}, and disorders \cite{Hanus2021thermal}. Among various approaches to calculating \gls{ltc} \cite{Gu2021Thermal}, \gls{md} simulation plays a unique role due to its versatility and its natural inclusion of the full lattice  anharmonicity. \gls{md} simulations are widely applicable in crystals, glasses \cite{Lee1991molecular}, and also liquids \cite{vogelsang1987thermal}. Two basic categories are commonly used, including \gls{emd} base on the Green-Kubo formalism \cite{Green1954markoff,kubo1957statistical} and \gls{nemd} based on Fourier's law of heat conduction. Notably, the \gls{hnemd} method, initially developed by Evans \cite{Evans1982homogeneous} for pairwise interactions and recently generalized to many-body interactions  \cite{Fan2019homogeneous}, offers great efficiency for \gls{ltc} calculations. However, the applicability and predictive power of \gls{md} simulations have long been limited by the availability and accuracy of empirical interatomic potentials.

A promising solution to this issue involves constructing \glspl{mlp} trained against reference  energies, forces, and virial stresses of diverse atomic structures calculated at the quantum mechanical level. Many \glspl{mlp} have been used for thermal conductivity modeling. Enabled by \glspl{mlp}, the \glspl{ltc} of many crystals with strong phonon anharmonicity or disorder have been successfully obtained through \gls{mlmd}, including e.g., amorphous GeTe \cite{Sosso2012Thermal}, SnSe \cite{Liu2021High-temperature}, PbTe \cite{Cheng2023Lattice}, metal-organic frameworks \cite{Ying2023Sub}, and PH$_4$AlBr$_4$ \cite{Du2023Low}. Moreover, with proper quantum corrections, quantitative agreement with experimental data has also been achieved for amorphous materials \cite{Wang2023Quantum-corrected,Liang2023Mechanisms} and liquid water \cite{Xu2023Accurate} over a wide range of temperatures. Despite these successes, previous works have also shown that for materials with relatively high \glspl{ltc}, such as CoSb$_3$ \cite{Korotaev2019Accessing} and \gls{csi} \cite{Qian2019Thermal}, the predicted \glspl{ltc} from \gls{mlmd} calculations are generally lower than the experimental values. To the best of our knowledge, this discrepancy remains to be systematically understood and corrected, which constitutes the main focus of our present work.

In light of the critical impact of the interatomic forces on the accuracy of \gls{md} simulations, we first evaluate the effect of random forces on \gls{ltc} using \gls{hnemd} simulations with a Langevin thermostat \cite{Bussi2007Accurate} based on empirical potentials. A decrease in \gls{ltc} with increasing level of random forces is consistently observed in six representative materials: \gls{asi}, \gls{csi}, \gls{cge}, Si-Ge alloy, graphene, and $(10, 10)$-\gls{cnt}.  Subsequently, we focus on four benchmark materials including \gls{csi}, GaAs, graphene, and PbTe, and perform \gls{ltc} calculations using \gls{mlmd}. In particular, we employ the \gls{nep} \cite{Fan2021Neuroevolution,Fan2022Improving,Fan2022GPUMD} for its balanced efficiency and accuracy. Similar to literature results, we  observe a consistent underestimation of \gls{ltc} from the \gls{mlmd} simulations, as compared to the experimental values. However, since the residual force errors of a NEP model and the random forces (white noises) in the Langevin thermostat both follow a Gaussian distribution, we propose an approach to correcting the \gls{ltc} by intentionally introducing different levels of force noises via the Langevin thermostat and then extrapolating to the limit of zero force error. This extrapolation successfully corrects the \glspl{ltc}, leading to excellent agreement with experimental data for all the materials considered in a wide range of temperatures. Spectral analyses reveal that the \gls{ltc} underestimation before the correction mainly originates from increased phonon scatterings at low frequencies caused by the force errors.

\section{Methods}

\subsection{Neuroevolution potential}

\subsubsection{The NEP formalism} 

In this section, we briefly review the \gls{nep} formalism \cite{Fan2021Neuroevolution,Fan2022Improving,Fan2022GPUMD}.
\gls{nep} uses a feedforward neural network to correlate a local descriptor with the site energy $U_i$ of atom $i$. In a single-hidden-layer neural network comprising $N_{\rm neu}$ hidden neurons, $U_i$ is expressed as: 
\begin{equation}
\label{equation:u_i}
U_{i} = \sum_{\mu=1}^{N_{\rm neu}} \omega_\mu^{(1)} \tanh\left(\sum_{\nu=1}^{N_{\rm des}}\omega_{\mu\nu}^{(0)}q_{\nu}^i-b_{\mu}^{(0)}\right)-b^{(1)}, 
\end{equation}
where $N_{\rm des}$ is the number of descriptor components,  $q_{\nu}^i$ is the $\nu$-th descriptor component of atom $i$, $\omega_{\mu\nu}^{(0)}$, $\omega_\mu^{(1)}$, $b_{\mu}^{(0)}$, and $b^{(1)}$ are the trainable parameters, and $\tanh(x)$ is the nonlinear activation function in the hidden layer.

The descriptor vector in \gls{nep} includes radial and angular components. The radial components $q_n^i $ $(0 \leq {n} \leq n_{\rm max}^{\rm R})$ are defined as
\begin{equation}
\label{equation:rad_des}
q_{n}^i = \sum_{j\neq{i}}g_n(r_{ij}), 
\end{equation}
where $r_{ij}$ is the distance between atoms $i$ and $j$ and $g_n(r_{ij})$ are a set of radial functions, each of which is formed by a linear combination of Chebyshev polynomials. The angular components include the so-called $n$-body ($n\geq 3$) correlations. For example, the $3$-body ones $q_{nl}^i$ $(0 \leq {n} \leq n_{\rm max}^{\rm A}$,  $1 \leq {l} \leq l_{\rm max})$ are defined as
\begin{equation}
\label{equation:ang_des}
q_{nl}^i =\frac{2l+1}{4\pi}\sum_{j\neq{i}}\sum_{k\neq{i}}g_n(r_{ij})g_n(r_{ik})P_l(\cos\theta_{ijk}).
\end{equation}
Here, $P_l$ is the Legendre polynomial and $\theta_{ijk}$ is the angle formed by the $ij$ and $ik$ bonds. Note that the radial functions $g_n(r_{ij})$ for the radial and angular descriptor components can have different cutoff radii, which are denoted as $r_{\rm c}^{\rm R}$ and $r_{\rm c}^{\rm A}$, respectively. The free parameters are optimized using the separable natural evolutionary strategy \cite{Schaul2011High} by minimizing a loss function that is a weighted sum of the \glspl{rmse} of energy, force, and virial stress, for $N_{\rm gen}$ generations with a population size of $N_{\rm pop}$. The hyperparameters used for all the materials considered in this work are listed in \textcolor{blue}{Table S1}.

\subsubsection{Training datasets}

For \gls{csi}, GaAs, graphene, and PbTe, we generate datasets through \gls{dft} calculations using the \textsc{vasp} \cite{Blochl1994Projector} with the ion-electron interactions described by the projector-augmented wave method \cite{Blochl1994Projector, Kress1999From}. For GaAs, the Perdew-Zunger functional with the local density approximation \cite{Perdew1981Self} is used to describe the exchange-correlation of electrons, while the Perdew-Burke-Ernzerhof functional with the generalized gradient approximation  
\cite{Perdew1997Generalized} is used for the other materials. 
The cutoff energy is 400 eV for PbTe and 600 eV for the other materials. The k-point mesh is $4\times4\times4$ for \gls{csi}, $2\times2\times2$ for GaAs and PbTe, and $6\times6\times1$ for graphene. The energy convergence threshold is $10^{-6}$ eV for \gls{csi} and graphene and $10^{-8}$ eV for GaAs and PbTe.

The dataset for each materials consists of structures from \gls{aimd} simulations (called \gls{aimd} structures below) possibly supplemented by those from random cell deformations and atom displacements (called perturbation structures below). For \gls{csi}, there are 900 \gls{aimd} structures sampled at various temperatures (100 K to 1000 K) and strain states (unstrained, uniaxial strains of $\pm1\%$ and $\pm2\%$, biaxial strains of $\pm0.5\%$ and $\pm1\%$) and 70 perturbation structures. Each \gls{csi} structure has 64 atoms. For GaAs, there are 197 \gls{aimd} structures sampled at various temperatures (100 K to 900 K) in the $NPT$ ensemble and 99 perturbation structures up to ±4\% strains. Each GaAs structure has 250 atoms. 
For graphene, there are 700 \gls{aimd} structures sampled at various temperatures (100 K to 1000 K) and strain states (unstrained, biaxial strains of $\pm0.5\%$, $\pm1\%$, and $2\%$). Each graphene structure has 72 atoms. For PbTe, there are 60 \gls{aimd} structures sampled from 100 to 1100 K with fixed cell and 64 perturbation structures up to ±4\% strains. Each PbTe structure has 216 atoms.  

After obtaining the datasets, we used the \textsc{gpumd} package \cite{Fan2017Efficient} (the \verb|nep| executable) to train the \gls{nep} models. The parity plots and accuracy metrics are detailed in \textcolor{blue}{Figs. S1-S4}. Force test errors will be further discussed and used in \autoref{force_noise}.

\subsection{Thermal conductivity calculation using MD}

\subsubsection{The HNEMD method} 

We use the efficient \gls{hnemd} method \cite{Fan2019homogeneous} for many-body potentials to calculate the \glspl{ltc}. 
In \gls{hnemd}, an external driving force on each atom $i$
\begin{equation}
\mathbf{F}_{i}^{\rm ext}=\mathbf{F}_{\rm e} \cdot \mathbf{W}_i    
\end{equation}
is applied during the simulation. Here, $\mathbf{F_{\rm e}}$ is the driving force parameter (of the dimension of inverse length) and \cite{Fan2021Neuroevolution,Fan2022GPUMD}
\begin{equation}
\label{equation:virial}
\mathbf{W}_i=\sum_{j \neq i}\mathbf{r}_{ij} \otimes \frac{\partial{U_j}}{\partial \mathbf{r}_{ji}}
\end{equation}
is the virial tensor of atom $i$, where $U_j$ is the site energy of atom $j$, $\mathbf{r}_{ij} \equiv \mathbf{r}_j - \mathbf{r}_i$, $\mathbf{r}_i$ being the position of atom $i$.
The driving force parameter should be large enough to ensure a large signal-to-noise ratio and be small enough to maintain the system in the linear-response regime. In the linear-response regime, the \gls{ltc} tensor $\kappa_{\mu\nu}$ can be calculated from the following relation \cite{Fan2019homogeneous}: 
\begin{equation}
\label{equation:hnemd}
\frac{\langle J_{\mu}(t)\rangle_{\rm ne}}{TV} = \sum_{\nu} \kappa_{\mu\nu}{F_{\rm e}}^\nu,
\end{equation}
where $\langle J_{\mu}(t)\rangle_{\rm ne}$ represents a non-equilibrium ensemble average of the heat current, $T$ is the system temperature, and $V$ is the system volume. The heat current for the \gls{nep} model has been derived to be \cite{Fan2021Neuroevolution,Fan2022GPUMD}
\begin{equation}
    \mathbf{J} = \sum_i \mathbf{W}_i \cdot \mathbf{v}_i,
\end{equation}
where $\mathbf{v}_i$ is the velocity of atom $i$.

The \gls{hnemd} formalism also allows for an efficient calculation of the frequency-resolved \gls{ltc} $\kappa(\omega)$ via the following relation \cite{Fan2019homogeneous}: 
\begin{equation}
\label{equation:shc1}
\frac{2}{VT} \int_{-\infty}^{+\infty} e^{i\omega t} K^{\mu}(t) dt= \sum_{\nu} \kappa_{\mu\nu}(\omega){F_{\rm e}}^\nu,
\end{equation}
where 
\begin{equation}
\label{equation:shc2}
K^{\mu}(t) = \sum_i \sum_{\nu}\langle W_i^{\mu\nu}(0) v_i^{\nu}(t)\rangle_{\rm ne}
\end{equation}
is the virial-velocity correlation function. 

\subsubsection{Thermostats in HNEMD simulations}

\gls{hnemd} simulations are normally performed in the $NVT$ ensemble realized by using a \textit{global} thermostat such as the \gls{nhc} \cite{Martyna1992Nosé} or the Bussi-Donadio-Parrinello \cite{Bussi2007Canonical} thermostat. In contrast, a \textit{local} thermostat such as the Langevin thermostat \cite{Bussi2007Accurate} is avoided because it can introduce (white) noises through random forces, leading to the following equations of motion:
\begin{equation}
\label{equation:langevin}
\frac{d\bm{\mathrm{r}}_i}{dt}=\frac{\bm{\mathrm{p}}_i}{m_i},\hspace {1cm}  \frac{d\bm{\mathrm{p}}_i}{dt}=\bm{\mathrm{F}}_i-\frac{\bm{\mathrm{p}}_i}{\tau_T} +\bm{\mathrm{f}}_i.
\end{equation}
Here, $\tau_T$ is a time parameter, $\mathbf{r}_i$, $\mathbf{p}_i$, $m_i$ are respectively the position, momentum, and mass of atom $i$, $\mathbf{F}_i$ is the force on atom $i$ resulting from the interatomic potential, and $\mathbf{f}_i$ is the random force on atom $i$. Each component of the random force forms a Gaussian distribution with zero mean and a variance of 
\begin{equation}
\label{equation:sigma_L}
    \sigma^2_{\rm L}= \frac{2k_{\rm B}Tm}{\tau_T \Delta t}, 
\end{equation}
where $m$ is the average atom mass in the system, $k_{\rm B}$ is the Boltzmann constant and $\Delta t$ is the integration time step. The random forces can affect the dynamics of the system and thus time-correlation properties such as the heat current autocorrelation function, leading to reduced \gls{ltc} as compared to the case of using a global thermostat. Clearly, a smaller $\tau_T$ gives a larger random force variance and a stronger reduction of the \gls{ltc}. We will demonstrate this effect using examples.

\subsubsection{MD simulation details}

All the \gls{md} simulations are performed using the \textsc{gpumd} package \cite{Fan2017Efficient} (the \verb|gpumd| executable), with a time steps of 1 fs.
For all the materials, we use a sufficiently large simulation cell to eliminate finite-size effects.
In \gls{md} simulations with empirical potentials, the simulation cells contain \num{32768} atoms for Si-Ge alloy, \gls{cge}, \gls{csi}, and \gls{asi}, \num{15416} atoms for graphene, and \num{16280} atoms for the $(10,10)$-\gls{cnt}. The \gls{asi} samples are prepared by employing a melt-quench-anneal process, first equilibrating at 2000 K for 10 ns, then quenching down to 300 K during 30 ns, and finally annealing at 300 K for 10 ns. In \gls{md} simulations with \gls{nep} models, the simulation cells contain \num{13824}, \num{8000}, \num{16000}, and \num{36000} atoms for \gls{csi}, GaAs, graphene, and PbTe, respectively. These cells have been tested to be large enough to eliminate the finite-size effects in \gls{hnemd} simulations (see \textcolor{blue}{Fig. S7}). For each material, we first equilibrate the system in the $NPT$ ensemble (with a target pressure of zero) for 2 ns and $NVT$ ensemble for another 2 ns, and then calculate the \gls{ltc} in the $NVT$ ensemble during a production time of 10 to 20 ns. For each material at each temperature, three to five independent runs are performed to improve the statistical accuracy and obtain an error estimate. The error bars are calculated from the statistical standard error of independent simulations. An example of c-Si at 300 K is shown in \textcolor{blue}{Fig. S6}.

\section{Results and discussion}

\subsection{Thermal conductivity underestimation in MLMD}

\begin{figure}[ht]
\centering
\includegraphics[width=0.8\linewidth]{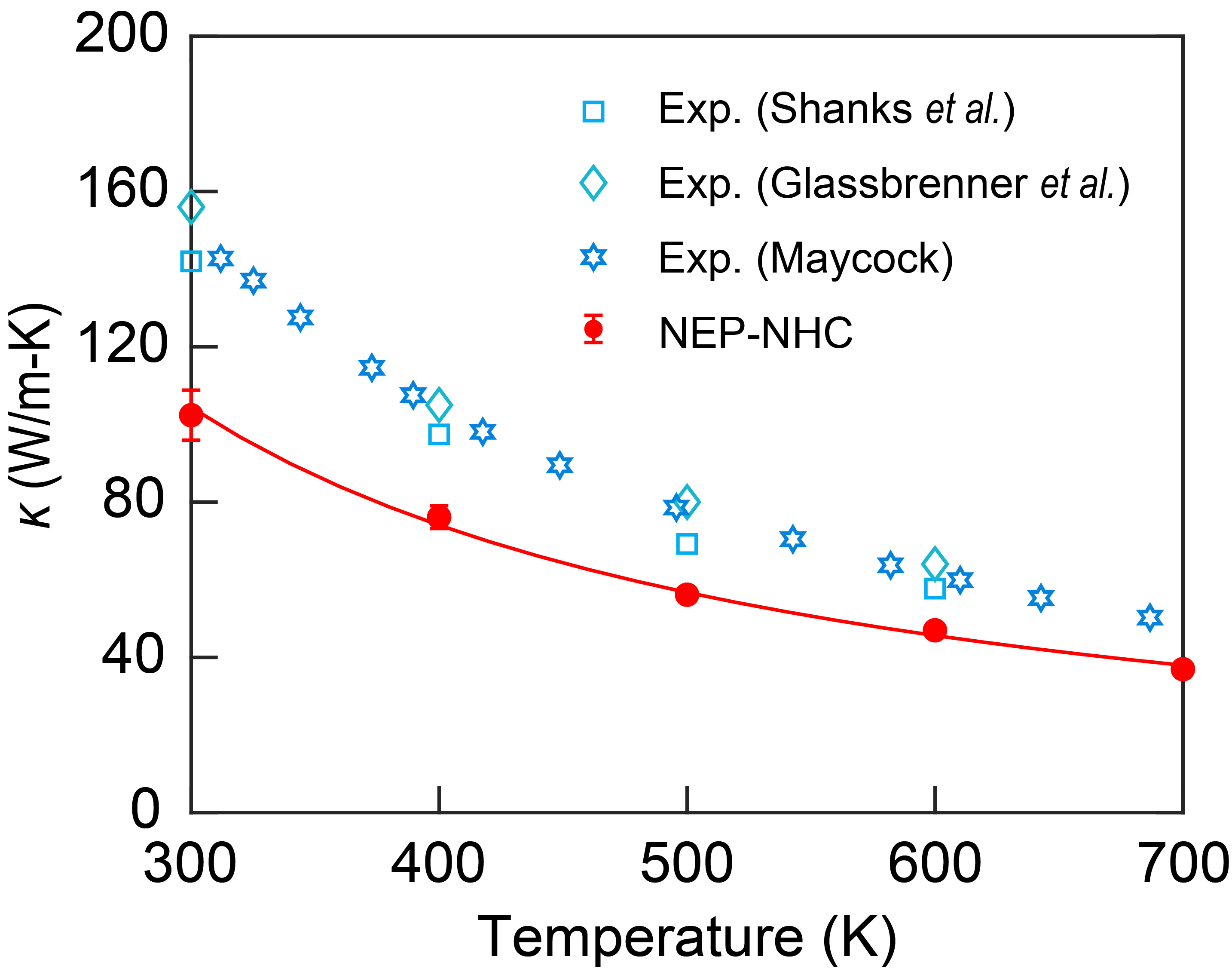}
\caption{Comparison of $\kappa$ for \gls{csi} from \gls{nep}-\gls{md} simulations and experimental measurements \cite{Glassbrenner1964Thermal,Shanks1963Thermal,Maycock1967Thermal}. Here, the \gls{nhc} thermostat is used. Error bars are smaller than the symbol sizes for the calculated values. }
\label{figure:Si-kappa}
\end{figure}

To begin with, we take \gls{csi} as an example to demonstrate the thermal conductivity underestimation in MLMD simulations, using \gls{nep} as a representative MLP. As illustrated in \autoref{figure:Si-kappa}, the calculated \gls{ltc} values from 300 K to 700 K are consistently lower than experimental measurements, especially at low temperatures. For instance, at 300 K, \gls{mlmd} simulations yield a \gls{ltc} of $102 \pm 6$ W/m-K. While this is more accurate than the value of 240 W/m-K as obtained from a Stillinger-Weber potential \cite{Volz2000Molecular-dynamics}, it is still approximately 32\% lower than the experimental value of about 150 W/m-K \cite{Glassbrenner1964Thermal}. An \gls{emd} simulation based on the Gaussian approximation potential (GAP) also reported a lower-than-experiment value of 121 W/m-K \cite{Qian2019Thermal}. A similar trend of underestimation is observed for \gls{csi} at other temperatures by GAP \cite{Qian2019Thermal}, for GeTe by NEP \cite{zhang2024thermal}, and for CoSb$_3$ \cite{Korotaev2019Accessing} by moment tensor potential.

\subsection{Role of force noises in reducing LTC\label{force_noise}}

\begin{table}[ht]
\small
  \begin{center}
    \caption{\glspl{rmse} $\sigma_{\rm mlp}$ of force prediction for the four \gls{nep} models at various temperatures.}
    \label{table:error_of_NEP}
    \tabcolsep=0.4 cm 
    \begin{tabular}{lllll} 
    \hline
    \hline
       \multirow{2}{*}{$T$ (K)} & \multicolumn{4}{c}{ $\sigma_{\rm mlp}$ (meV/\AA)}\\
       \cline{2-5}
       & \gls{csi} & GaAs &graphene & PbTe\\
      \hline
      300 & 16.7 & 16.6 & 29.2 & 27.0\\ 
      400 & 21.3 & 19.8 & 30.1 & 29.9\\ 
      500 & 28.3 & 23.5 & 32.5 & 34.4\\ 
      600 & 30.1 & 26.8 & 36.6 & 37.1\\ 
      700 & 41.6 & 30.2 & 42.4 & 42.1\\ 
      \hline
      \hline
    \end{tabular}
  \end{center}
\end{table} 

\begin{figure}[ht]
\centering
\includegraphics[width=0.8\linewidth]{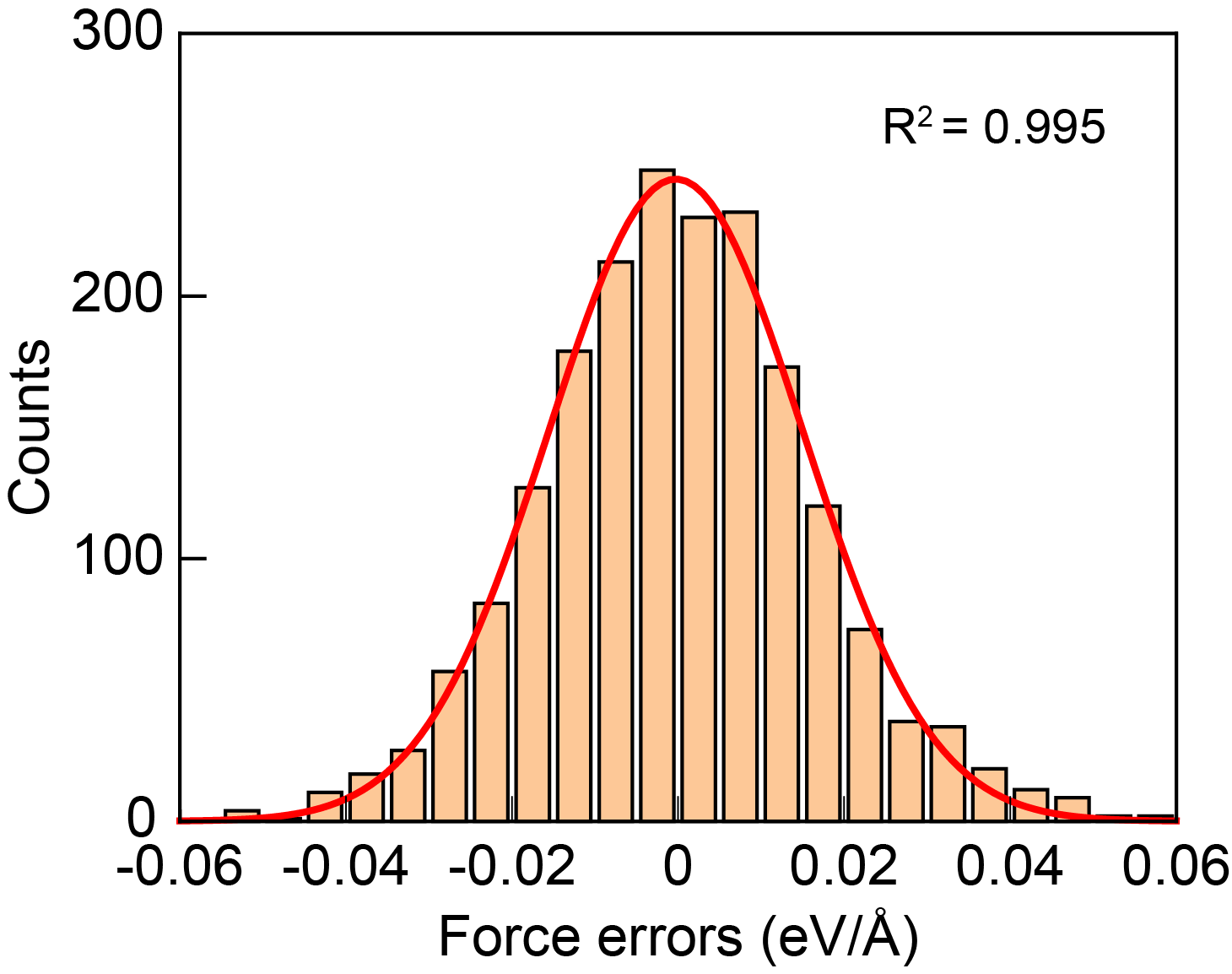}
\caption{Force error distribution for \gls{csi} at \emph{T} = 300 K. Fitting to the Gaussian distribution yields a coefficient of determination $R^2=0.995$.}
\label{figure:force-distri}
\end{figure}

\begin{figure*}[ht]
\centering
\includegraphics[width=0.97\linewidth]{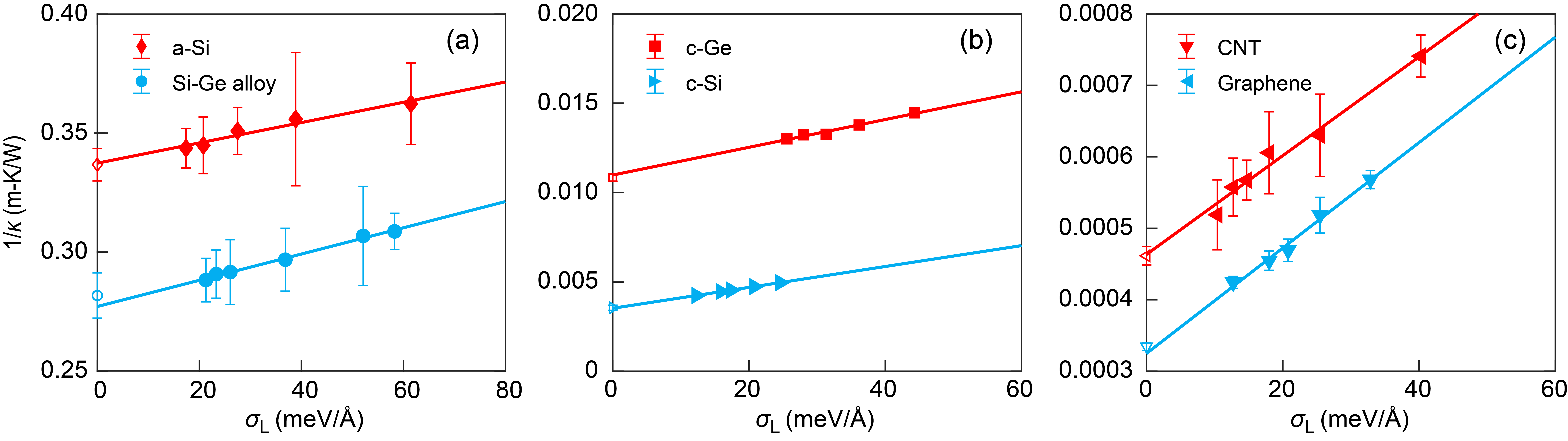}
\caption{Inverse \gls{ltc} ($1/\kappa$) as a function of the random force variance $\sigma_{\rm L}$ of the Langevin thermostat (see \autoref{equation:sigma_L}) for (a) \gls{asi} and Si-Ge alloy, (b) c-Ge and \gls{csi}, and (c) graphene and $(10,10)$-\gls{cnt} at 300 K. The hollow and filled symbols are the results from the NHC and Langevin thermostats, respectively. The solid lines represent linear fits to the Langevin data only. }
\label{figure:classicalMD}
\end{figure*}

\begin{figure*}[ht]
\centering
\includegraphics[width=0.8\textwidth]{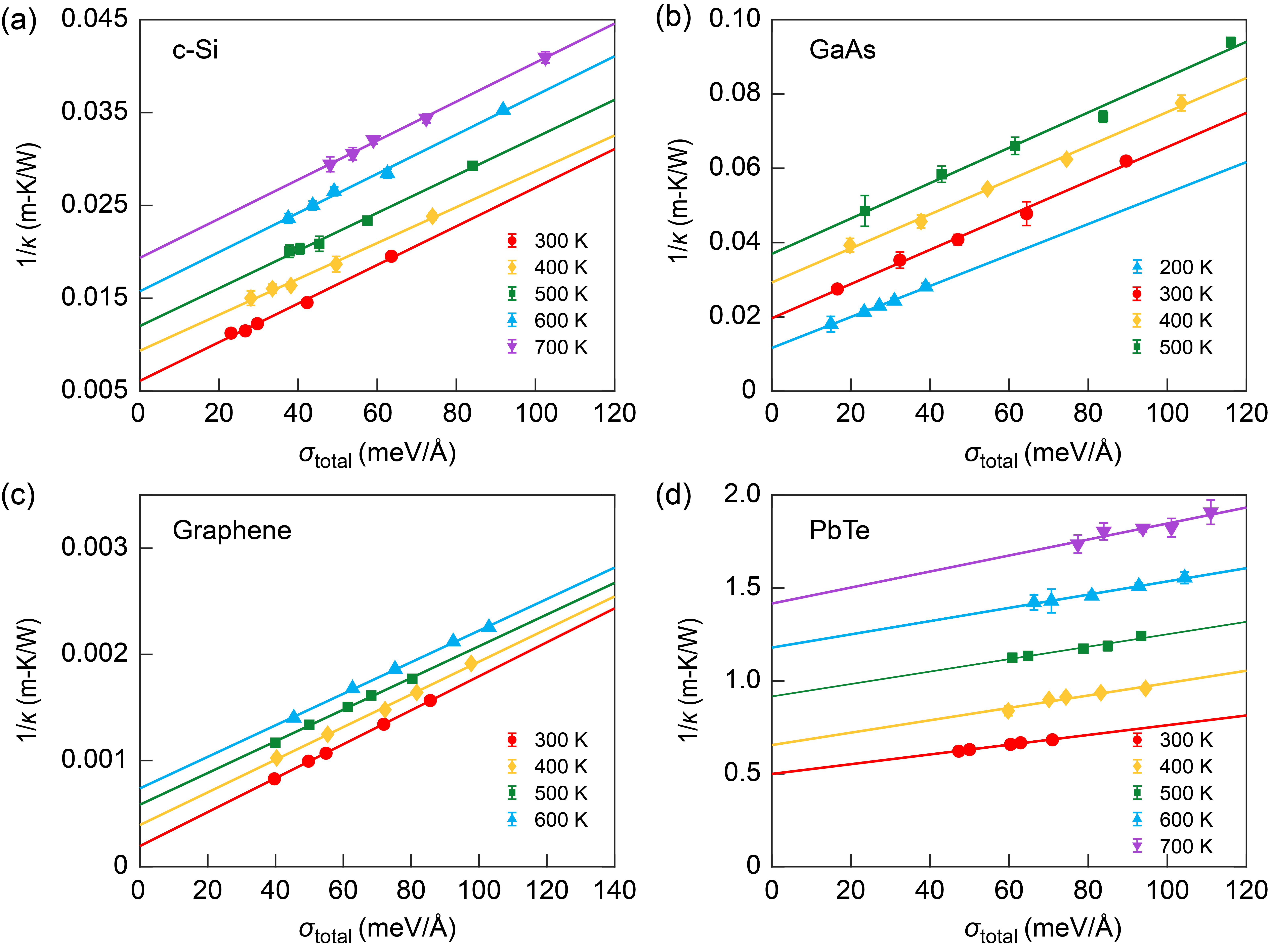}
\caption{Inverse LTC ($1/\kappa$) from \gls{nep}-\gls{md} simulations as functions of the total force error $\sigma_{\rm total}$ at different temperatures for (a) \gls{csi}, (b) GaAs, (c) graphene, and (d) PbTe. Solid lines indicate linear fits and the points of intersection at $\sigma_{\rm total}=0$ correspond to the corrected \gls{ltc} values.}
\label{figure:NEP_error_on_kappa}
\end{figure*}

To understand the underestimation of the \gls{ltc} from \gls{mlmd} simulations, we notice that a \gls{mlp} usually has a certain level of error for force prediction compared to the reference data. The \glspl{rmse} $\sigma_{\rm mlp}$ of force prediction for the four materials we considered at different temperatures are presented in \autoref{table:error_of_NEP}.

A crucial observation is that the force errors follow a Gaussian distribution, as shown in \autoref{figure:force-distri} for the example of \gls{csi} at 300 K. This distribution is the same as that for the random forces in the Langevin thermostat, i.e., a Gaussian distribution with zero mean and a certain variance. Based on this similarity, an understanding of the underestimation of the \gls{ltc} by \gls{mlmd} simulations can thus be obtained by studying the effect of the Langevin thermostat on the \gls{ltc}. When the system is coupled to the Langevin thermostat, a random frictional force will be added to all atoms, affecting the dynamics of the system. According to the Newton's equation of motion, the effect of random forces on the atoms is similar to that from randomly varied atomic masses. Thus, the coupling to Langevin thermostat introduces an additional phonon scattering term, the strength of which can be tuned by varying the coupling constant. One could directly use a \gls{nep} model for this test, but due to the lower computational cost of empirical potentials, we first use the Tersoff empirical potential \cite{tersoff1988Empirical} to study this effect.

\begin{figure*}[!hbt]
\centering
\includegraphics[width=0.8\textwidth]{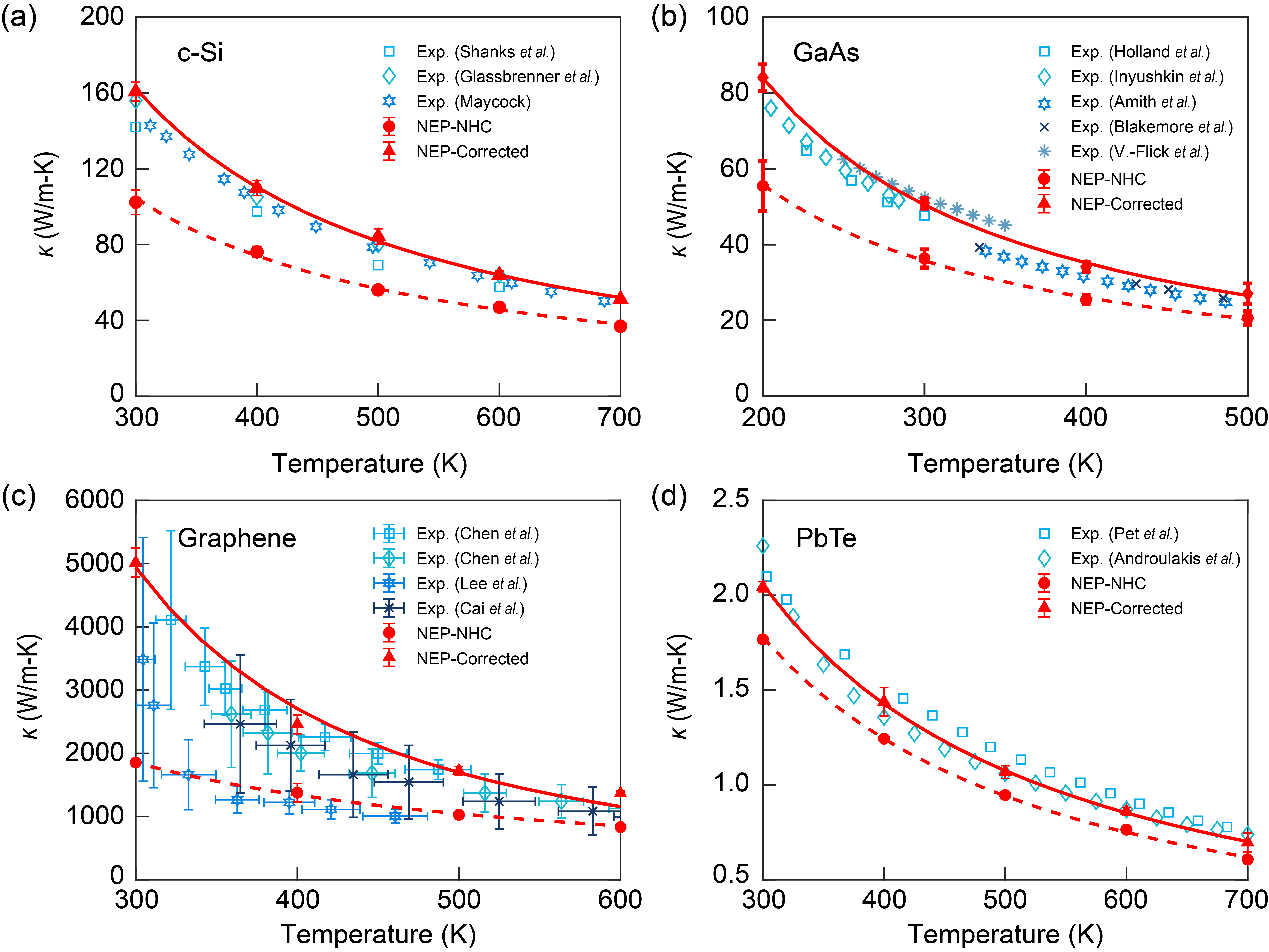}
\caption{\label{} Corrected (using the Langevin thermostat and exptrapolation) and uncorrected (using the \gls{nhc} thermostat) $\kappa$ as a function of temperature for (a) \gls{csi}, (b) GaAs, (c) graphene, and (d) PbTe. Experimental values are from Ref. \cite{Glassbrenner1964Thermal, Shanks1963Thermal, Maycock1967Thermal} (\gls{csi}), Ref. \cite{Amith1965Electron,Blakemore1982Semiconducting,Vega-Flick2019Reduced,Inyushkin2003Thermal,Holland1964Phonon} (GaAs), Ref. \cite{Chen2011Raman,Chen2012Thermal,Lee2011Thermal,Cai2010Thermal} (graphene), and Ref. \cite{Androulakis2010Thermoelectric, Pei2011Convergence} (PbTe). It is worth to note that the experimentally synthesized samples may contain defects such as vacancies and dislocations. Besides, the synthesized samples usually have limited sizes. Thus the experimentally measured samples may involve weak defect and boundary scatterings, leading to slight deviations between measured and predicted thermal conductivities.}
\label{figure:correct}
\end{figure*}

In \autoref{figure:classicalMD}, we show the inverse \gls{ltc} ($1/\kappa$) at 300 K as a function of $\sigma_{\rm L}$ for six representative materials, including \gls{asi}, Si-Ge alloy, \gls{cge}, \gls{csi}, graphene, and $(10,10)$-CNT. As expected, $1/\kappa$  increases with increasing $\sigma_{\rm L}$, which indicates a stronger effect of the random forces in the Langevin thermostat in reducing the calculated \gls{ltc}. Notably, for all the six materials, $1/\kappa$ exhibits a linear relationship with $\sigma_{\rm L}$. This suggests that the intrinsic \gls{ltc} without the influence of the random forces in the Langevin thermostat can be obtained by extrapolating to $\sigma_{\rm L}=0$. Indeed, the extrapolated values align well with the results from \gls{hnemd} simulations based on the \gls{nhc} thermostat that does not involve random forces, with the largest relative error being $<1.5\%$ (see \textcolor{blue}{Table S2}). 

The linear relation between $1/\kappa$ and $\sigma_{\rm L}$ can be justified based on the kinetic theory of phonons and Matthiessen's rule. Taking the random forces in the Langevin thermostat as an extra source of phonon scattering, we have
\begin{equation}
\label{equation:Mat_rule}
\frac{1}{\kappa}=\frac{1}{\kappa_0} + \frac{1}{1/3Cv_{\rm g}\Lambda_{\rm L}},
\end{equation}

where $\kappa$ and $\kappa_0$ are the \glspl{ltc} with and without the influence of the random forces, respectively, $C$ is the heat capacity, $v_{\rm g}$ is the phonon group velocity, and $\Lambda_{\rm L}$ is the phonon mean free path resulting from the random forces in the Langevin thermostat. Under first-order approximation with sufficiently small $\sigma_{\rm L}$, $1/\Lambda_{\rm L}$ should be proportional to $\sigma_{\rm L}$, which brings \autoref{equation:Mat_rule} to

\begin{equation}
\label{equation:Mat_rule2}
\frac{1}{\kappa}=\frac{1}{\kappa_0} + \beta \sigma_{\rm L},
\end{equation}

\noindent which gives the observed linear relation between $1/\kappa$ and $\sigma_{\rm L}$ with $\beta$ being a slope parameter.

\subsection{Correction of LTC in MLMD}

Based on the results above, we can understand why \gls{mlmd} usually underestimates the \gls{ltc}, particularly for high-$\kappa$ materials. According to the linear relation between the inverse \gls{ltc} and the random force variance, we can devise a method to correct the underestimation of \gls{ltc} due to the force errors in \gls{mlmd}. To this end, we note that both the force errors in \gls{mlmd} and the random forces in the Langevin thermostat follow a Gaussian distribution, and when they are present simultaneously, a new set of force errors are created with a larger variance given by
\begin{equation}\label{equation:sum_sigma}
    {\sigma_{\rm total}}^2={\sigma_{\rm L}}^2+{\sigma_{\rm mlp}}^2,
\end{equation}
according to the properties of Gaussian distribution. Therefore, we can intentionally introduce extra force errors by using MLP-based \gls{hnemd} simulations with the Langevin thermostat. The \gls{ltc} $\kappa_0$ without any force errors (including the force errors of the \gls{mlp}) can be obtained by an extrapolation based on the following relation:
\begin{equation}
\label{equation:kappa_no_error}
\frac{1}{\kappa} =\frac{1}{\kappa_0} + \beta \sigma_{\rm total},
\end{equation}
where $\kappa$ is the \gls{ltc} of a material calculated by using \gls{mlmd} with a certain force error variance $\sigma_{\rm mlp}$ and the Lagevin thermostat with a certain random force variance $\sigma_{\rm L}$.
The linear relation between $1/\kappa$ and $\sigma_{\rm total}$ is unfailingly confirmed in \autoref{figure:NEP_error_on_kappa} for the four representative materials in a wide range of temperatures, whose \glspl{ltc} span three orders of magnitude.

In \autoref{figure:correct}, we compare the uncorrected and corrected \glspl{ltc} from \gls{mlmd} simulations with experimental results for \gls{csi}, GaAs, graphene, and PbTe. In all the systems, the uncorrected \glspl{ltc} are consistently lower than the experimental results in the entire temperature range due to the presence of force errors in the \glspl{mlp}. Remarkably, once the force errors in the \glspl{mlp} are eliminated via our extrapolation scheme, the \glspl{ltc} closely approach the experimental data at all the temperatures studied. For graphene, the corrected \glspl{ltc} slightly exceed the measured values but remain within the experimental uncertainties. This minor discrepancy could arise from factors such as isotope scattering and finite-size effects in the experimental setups \cite{Chen2011Raman,Chen2012Thermal,Lee2011Thermal,Cai2010Thermal}, which generally lead to reduced \glspl{ltc}.

\begin{figure}[ht]
\centering
\includegraphics[width=0.8\linewidth]{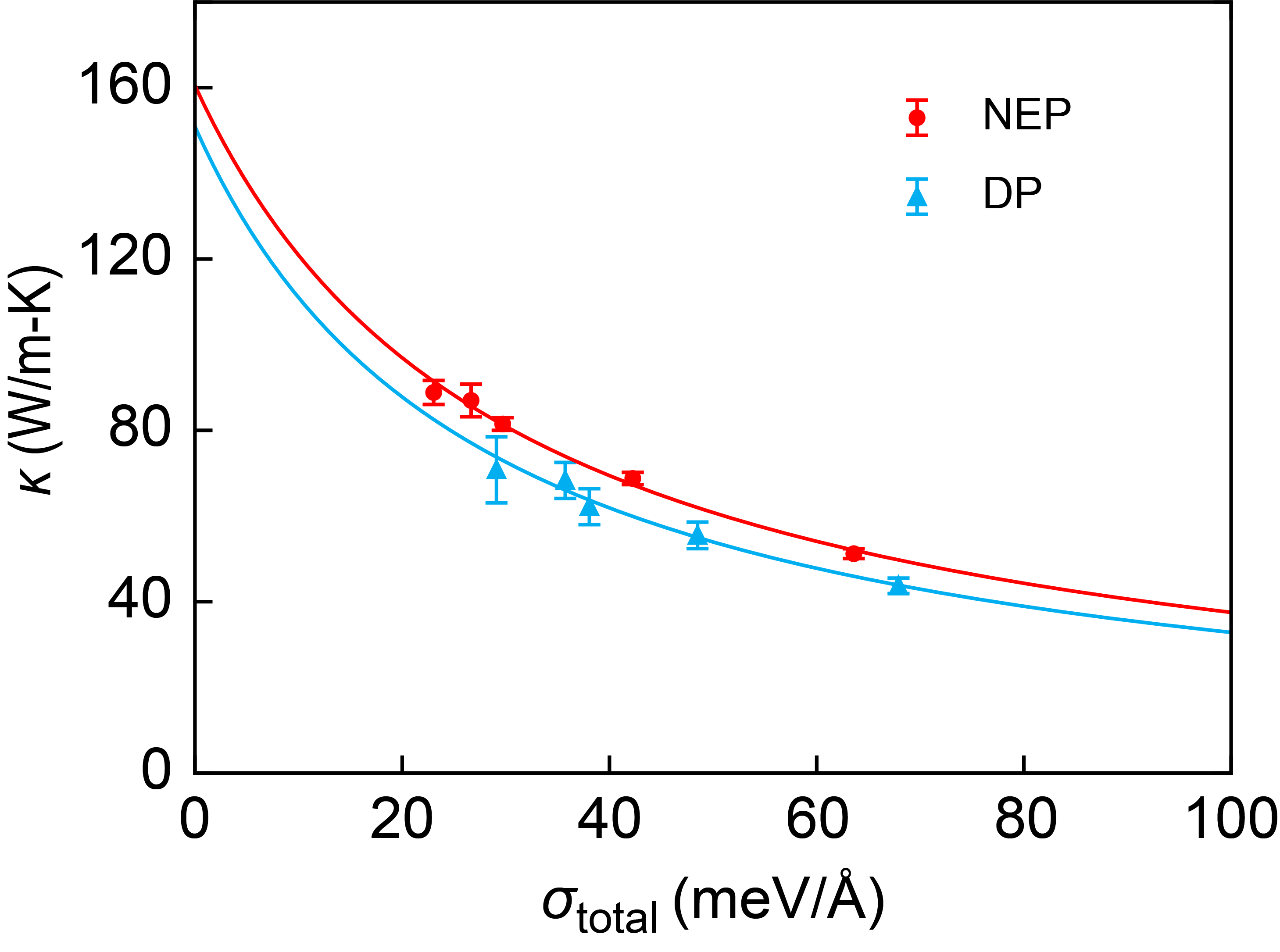}
\caption{ Calculated LTC for c-Si from NEP-HNEMD and DP-EMD simulations as a function of the total force error $\sigma_{\rm total}$. The DP-EMD results are obtained from 20 independent runs, each with a production time 2 ns.}
\label{figure:DP-NEP}
\end{figure}
 
\begin{figure}[ht]
\centering
\includegraphics[width=0.8\linewidth]{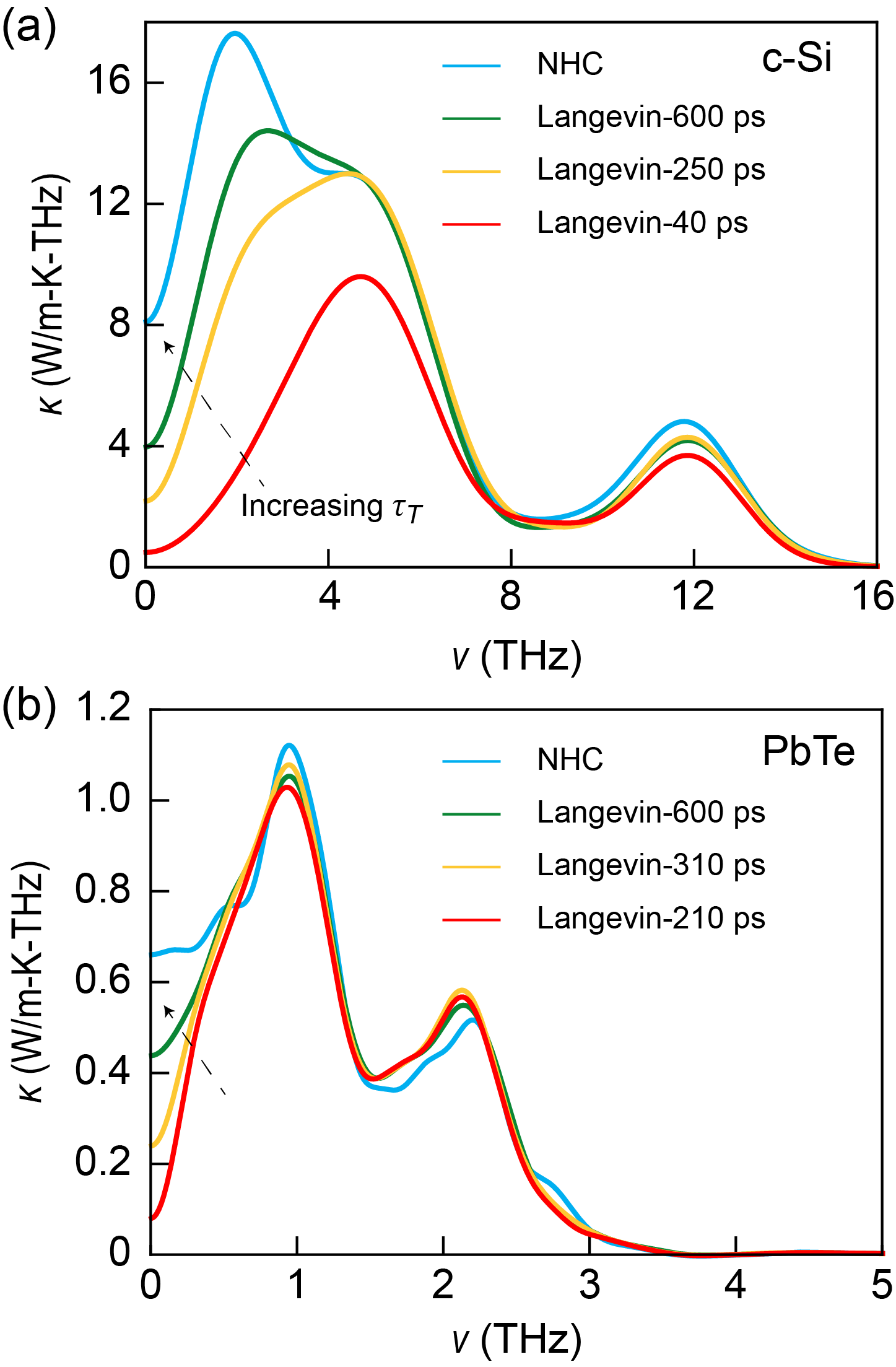}
\caption{Spectral \gls{ltc} $\kappa(\omega)$ from \gls{nep}-\gls{md} simulations using the \gls{nhc} and Langevin thermostats for (a) \gls{csi} and (b) PbTe, both at 300 K.
}
\label{figure:spectral}
\end{figure}

To demonstrate that the thermal conductivity underestimation is not specific to \gls{nep}, we consider the \gls{dp} \cite{zhang2018deepl, wang2018deepmd} as an additional example.  We train a \gls{dp} model for silicon using the same training dataset as used for NEP. The force RMSE at 300 K is determined to be 29.0 meV/\AA{}. Because \gls{hnemd} is not available for the \gls{dp} model via the LAMMPS MD engine \cite{plimpton1995fast}, we perform \gls{emd} simulations instead. Similar to \gls{hnemd} simulations, we use the Langevin thermostat with coupling times of 350, 250, 100, and 40 ps to introduce additional force errors, giving rise to total force errors of 35.7, 38.0, 48.4, and 67.9 meV/\AA, respectively. We also use the \gls{nhc} thermostat corresponding to the total force error of 29.0 meV/\AA. The results are shown in \autoref{figure:DP-NEP}. Clearly, the thermal conductivity predicted from the \gls{dp} model with no additional force errors (corresponds to the case of using the \gls{nhc} thermostat) is also underestimated compared to the experimental value. With the decrease of coupling time in the Langevin thermostat, the thermal conductivity reduces gradually. Based our proposed extrapolation formula  \autoref{equation:kappa_no_error}, the corrected thermal conductivity from \gls{dp} is 151 W/m-K, which is very close to the one obtained by NEP (160 W/m-K) with a relative difference of $\sim 5\%$. Therefore, we conclude that the underestimation of \gls{ltc} is a common issue in \glspl{mlp} and can be corrected by our proposed method.

The need for \gls{ltc} correction is more pronounced in materials with higher \glspl{ltc} and at lower temperatures. This is attributed to the weaker anharmonic phonon-phonon interactions, which leads to a relatively stronger contribution of the phonon scattering by the force errors.
This also explains why the amount of correction is large for graphene that is one of the most thermally conductive material, intermediate for \gls{csi} and GaAs that have intermediate \glspl{ltc}, and small for PbTe that has low \gls{ltc}. Furthermore, the spectral \gls{ltc} results in \autoref{figure:spectral} show that the force errors mainly reduce $\kappa(\omega)$ in the low-frequency region. With increasing force errors, $\kappa(\omega)$ in the low-frequency region is more and more reduced. This further supports the large effect of the force errors in high-\gls{ltc} materials, which usually have large $\kappa(\omega)$ in the low-frequency region. Therefore, \gls{mlmd} simulations remain largely accurate for low-\gls{ltc} materials, such as PbTe \cite{Cheng2023Lattice}, \gls{asi} \cite{Wang2023Quantum-corrected}, amorphous SiO$_2$ \cite{Liang2023Mechanisms}, and liquid water \cite{Xu2023Accurate}.

\section{Conclusions}

In summary, our systematic investigation revealed that the underestimation of lattice thermal conductivity commonly observed in the literature are primarily due to force fitting errors in machine learned potentials. Using empirical potentials and Langevin thermostat we demonstrated that introducing random forces on atoms can significantly reduce the lattice thermal conductivity, supporting our hypothesis. These random forces act as an additional source of phonon scattering, thereby reducing the lattice thermal conductivity. Employing the kinetic theory of phonons and Matthiessen's rule, we established a linear extrapolation formula to estimate the thermal conductivity in the absence of random forces. The validity of the extrapolation scheme was tested using empirical potentials on various materials, including \gls{asi}, Si-Ge alloys, \gls{csi}, \gls{cge}, graphene, and \gls{cnt}. 

We established that the force errors in machine-learned potentials follow a Gaussian distribution, akin to the distribution of random forces in the Langevin thermostat. This similarity inspired us to intentionally introduce extra force noises via the Langevin thermostat and then extrapolate to the limit of zero force error. The extrapolated results show excellent agreement with experimental data over a broad temperature range for all the materials studied. Spectral thermal conductivity analyses further indicate that the underestimation of the lattice thermal conductivity is mainly due to increased acoustic phonon scatterings caused by the force errors. Our findings provide a clear explanation for the underestimated thermal conductivity often observed in molecular dynamics simulations based on machine learned potentials. The method of correcting this underestimation we developed will significantly enhance the applicability of machine learned potentials in the prediction of lattice thermal conductivity.

\vspace{0.5cm}
\noindent \textbf{SUPPLEMENTARY MATERIAL}
\vspace{1em}

See the supplementary material for the Hyperparameters used in NEP, the calculated thermal conductivity using empirical potentials, phonon dispersion relations from NEP models as compared to DFT calculations, size-convergence tests for thermal conductivity calculations, and the parity plots of trained NEPs.
%\end{SUPPLEMENTARY MATERIAL}

\vspace{0.5cm}
\noindent \textbf{Data availability}
\vspace{1em}

All the training datasets and the trained \glspl{nep} models are freely available at \url{https://gitlab.com/brucefan1983/nep-data}. The \gls{dp} training and potential files as well as \gls{md} input files are freely available at \url{https://github.com/hityingph/supporting-info}.

\begin{acknowledgments}
X.W. and W.Z. contributed equally. This work is supported by the National Natural Science Foundation of China (Grant No.12174276 and No. 52076002), the Basic and Applied Basic Research Foundation of Guangdong Province (Grant No. 2024A1515010521), the New Cornerstone Science Foundation through the XPLORER PRIZE. The Center of Campus Network and Modern Educational Technology of Guangdong University of Technology, and the High-performance Computing Platform of Peking University are acknowledged for providing computational resources and technical support for this work. P.Y. is supported by the Israel Academy of Sciences and Humanities \& Council for Higher Education Excellence Fellowship Program for International Postdoctoral Researchers.
\end{acknowledgments}

\vspace{0.5cm}
\noindent \textbf{Conflict of Interest}
\vspace{1em}

The authors have no conflicts to disclose.

%\bibliography{ref}

%aipnum4-2.bst 2019-01-14 (MD) hand-edited version of apsrev4-1.bst

\end{document}